# Independence of events and quantum structure of the light in Einstein's special relativity

Faculté des Sciences, Physique des particules, Université Libre de Bruxelles,
CP 230, Boulevard du Triomphe, ypiersea@ulb.ac.be
Subfaculty of Philosophy of Physics, University of Oxford, 10 Merton Street,
yves.pierseaux@philosophy.ox.ac.uk, 0044 1865 276933.
Fondation Wiener-Anspach

**1- Einstein's principles of relativistic kinematics**

**2- Einstein's 1905 proclamation of independence of events**

**3- Events, points of light and quanta of light in January paper**

**4- Events, points of light and complex of light in June paper**

**5- Poincaré's principles of implicit relativistic kinematics of deformable rods**

**6- Independent true time, material points and wave of light in Poincaré's SR with ether**

**Conclusion**

## 1- Einstein's principles of relativistic kinematics

Everything seems have been said about the famous miraculous year 1905 of the young Einstein. Many authors, physicists, philosophers and historians, have analysed the multiple connections between the three mains papers: quanta of light (January) Brownian motion (March) and special relativity (SR, June)[1].

The most important is perhaps this one of Louis de Broglie whose the source of inspiration for wavy mechanics is the contrast between the transformation of the light-frequency (Einstein January) and the transformation of the clock-frequency (Einstein June). Anyway it is well known that Einstein himself has developed a synthesis between quanta of light and Brownian motion in 1909 on duality of the structure of the radiation.

However, according also to Einstein himself, the three papers have not the same statute: the first and the second are "constructive theories" while the last – special relativity (SR) - is essentially a theory of principle [1].

January and March 1905 articles consist essentially in works of statistical thermodynamics that show the reality of atomic structure of matter (March) and the hypothesis of atomic structure of radiation (January).

Einstein's SR (June) looks deeply different because it is based on principles (more precisely two principles) that are independent of any consideration on the structure of the matter or radiation.

But, in Einstein's own words, thermodynamics and relativity are the best examples of a theory of principles. Einstein insists often on the analogy between the principles of special relativity and the principles of thermodynamics. In his response to Ehrenfest, about the question of rigidity, Einstein underlines even more precisely that "the principle of relativity together with the principle of constancy of velocity of light" must be conceived as a "heuristic principle" and also as being "similar to the second principle of thermodynamics".

We want to analyse Einstein's analogy in depth not to contest that Einstein's kinematics is based on principles (on the contrary!) but because there exists others principles in physics for example principle of reaction, principle of least action etc…(see for instance, Poincaré 1904 conference in Saint Louis on "principles of mathematical physics").

## 2- Einstein's 1905 proclamation of independence of events

We suggest here a new way of research because we can find in each of three Einstein's articles the concept of "independent events" (unabhängig Ereignis):

Quanta of light (January, paragraph 5)

In calculating entropy by molecular-theoretical methods, the word probability is often used in a sense differing from the way the word is defined in probability theory. In particular, case of equal probability are often hypothetically stipulated when the theoretical model employed are definite enough to permit a deduction rather than a stipulation. I will show in a separate paper that in dealing with thermal processes, it suffices completely to use the so-called *statistical probability* and hope thereby to eliminate a *logical difficulty* that still obstructs the application of *Boltzmann's principle*. Here, I shall give only its general formulation and its application to very special cases.
It is meaningful to talk about the *probability of a state* of a system, and if furthermore, each *increase of entropy* can be conceived as a transition to a state of higher probability, then the entropy $S_1$ of a system is a function of the probability of its *instantaneous state*. Therefore if we have two systems $S_1$ and $S_2$ that do not interact which each other, we can set $S_1 = \varphi (W1)$  $S_2 = \varphi (W2)$

It these two systems are viewed as a single system of entropy S and probability W, then:
$S = S_1 + S_2 = \varphi (W)$          and       $W = W_1 W_2$
The last equation tells us that the **states** of the two **systems are events that are independent of each other**(voneinander unabhängig Ereignis)". From these equations it follows that:
$$\varphi (W_1 . W_2) = \varphi (W_1) + \varphi (W_2)$$
If $S_0$ denotes the entropy of a certain initial state and W is the relative probability of a state of entropy S, we obtain [for Einstein always reference 1, always my italics and my bold]:

$$S - S_0 = k \ln W$$

Brownian motion (March, paragraph 4)

Obviously we must assume that each individual particle executes a motion that is independent of the motions of all the other particles; the motions of the same particle in different time intervals must also be considered as *mutually independent processes*, so long as we think of these time intervals as chosen not to be too small.
We now introduce a time interval τ, which is very small compared with observable time intervals but still large enough that the motions performed by a particle during two consecutive time intervals can be considered as **mutually independent events** (unabhängig Ereignis).[1]

Special relativity (June, paragraph 1)

We have to bear in mind that our judgements involving time are always judgments about **simultaneous events** (gleichzeitige Ereignisse). If, for example, I say that "the train arrives here at 7 o'clock" that means " the pointing of the small hand of my watch to 7 and the arrival of the train are **simultaneous events**".

We notice that in the three cases the concept of independent events is in the core of the article.

1) Quanta of light

Annus Mirabilis begins with Einstein's inversion of Boltzmann's principle: independent states or independent events are directly connected with Einstein's conception of statistical probability in one of the two most important paragraphs of the article (see3). I showed in my book [2] that Einstein's definition of probability with the inversion of Boltzmann's principle is remained essentially the same from 1903 to 1925 (gas of photons). It is interesting to notice about Einstein's logic that Einstein underlines the necessity "to eliminate a logical difficulty that still obstructs the application of Boltzmann's principle".

2) Brownian motion

Einstein treats fundamentally the Brownian motion as a random walk and the independent events take place in the most important paragraph of the paper where Einstein deduces his famous relation of fluctuation-diffusion of Brownian particle with which Perrin will show experimentally the reality of the atoms [2].

3) Special relativity

Einstein's simultaneous events at the same place are of course independent events. There is indeed no causal connection between the motion of the train that arrives in a given place (for example the station of Schaerbeek) and the motion of the small hand that arrives in a given place (for example the figure 7).

We showed [2] that the crucial importance of independence of events is not only true for the events at the same place (in Minkowki's cone of light) but also for distance events (on and out Minkowki's cone of light).

The reader could be have the impression that we play on the word "unabhangig Ereignisse" because in the two first cases the meaning of the word would be purely mathematical while only in the last case it would have a physical meaning connected with space and time.

We shall show in the details (this is the reason of the length of the quotations) that it is in the three case the same physical concept.

It is obvious for the *random walk* in the Brownian motion where each displacement D of the particle in a time t is an event. We insist on the fact it is the first time in the history of physics that the movement of a particle is described *as a series of independent events*.

In the article on quanta of light the deep physical meaning of Einstein's logic is directly visible in Einstein's identification of the concept of event with the concept of state ("the states of the two systems are events that are independent of each other"). Einstein's definition of independent states or events is also directly connected with the deduction of independent quanta of light (see 3).

The two first articles (January and March) establish clearly a discontinuous structure of matter and light. The standard look of Einstein's SR is, on the contrary, essentially based on the continuous conception of the field. The young Einstein however never uses in 1905 the concept of field while that he uses, all the time, the concept of event.

The advent of an event means deeply that "suddenly something happens somewhere" et therefore implies a concept of *discontinuity*. The collision is of course the best example of an event.

The classical continuous trajectory of a point material can be represented by a series of events but it is in this case "wrong" events or events in a weak sense. Indeed each event is completely determined by the previous event because the initial state x, v determines completely the next positions and velocities of the material point.

The "true" event (in the strong meaning of Prigogine) is thus a coincidence of events. This is precisely the definition given by Einstein in his famous paragraph on simultaneity in SR. We notice only about the Brownian motion that (see Einstein's quotation) the mutually independent *process* and mutually independent *event* are deeply connected in Einstein's logic because this connection is also valid for SR. So Einstein defines in his second synthesis on SR an event as the limit of a process ("To determine the position of a process of infinitesimally short duration that occurs in a space element (point event) we need a Cartesian system").

We want now to concentrate the attention on the synthesis between quanta of light and SR.

**3- Events, points of light and quanta of light in January paper**

Einstein characterises the *state of a system,* according to Boltzmann's principle, by the entropy S or the probability W. After his definition of probability of independent state or event, he considers a set (or a gas) of "points" (I use inverted comas because these points are not the usual material points of classical mechanics).

We now treat the following special case. Let a volume $V_0$ contain a number n of moving points (e.g. molecules) to which our discussion will apply. The volume may also contain any arbitrary number of other moving *points of any kind. No assumption is made about* **the law** *governing the motion* of the points under discussion in the volume except that, as concerns this motion, no part of the space (and no direction within it) is preferred over the others. Further, let the number of (aforementioned) moving points under discussion be small enough that we can disregard interactions between them.
This system, which, for example, can be an ideal gas or a dilute solution, possesses a certain entropy $S_0$. Let us imagine that this state has a different value of entropy (S), and we now wish to determine the difference in entropy with the help of **Boltzmann's principle**.

This space of points is homogenous and isotropic. Einstein applies therefore - in order to find the temporal evolution of the system ("each *increase of entropy* can be conceived as a transition to a state of higher probability") - not the second principle of Newton's mechanics but *the second principle of Boltzmann's thermodynamics*.
In the point of view of mechanics, Newton's law is the law of change of state – defined by the position x or q and velocity v or p- of the material point. In the point of view of the young Einstein, Boltzmann's law is the law of change of state[1] - defined by V and S - of the n independently moving "points":

We ask: how great is the probability of the last-mentioned state relative to the original one? Or: how great is the probability that a randomly chosen instant of time, all n **independently moving points** in a given volume $V_0$ will be found (by chance) in the volume V? Obviously, this probability, which is a "statistical probability" has the value:

$$W = (\frac{V}{V_0})^n$$

From this, by applying Boltzmann's principle, one obtains

$$S - S_0 = nk \ln \frac{V}{V_0}$$

Lorentz insisted, in Solvay Conference in 1911 [2], on the fact that Einstein's statistical probability has nothing to do with Gibbs's definition of probability based on the element dpdq in the space of phase. Anyway if the states of Einstein's points were defined by classical mechanics (in the sense of Laplace), it is impossible that the motions of the points were fundamentally independently. This is the deep reason for which Einstein rejects the definition of equiprobability (see quotation) based on "a mechanical priori model". According to Einstein, this a priori model is precisely this one of standard Gibbs's statistical mechanics.
Einstein's independently moving "points" suppose another definition of the state of these points. Einstein's thermodynamical definition (V, S) of state implies the definition of independent state not from a element of the phase space but from an "element of volume" and an "element of duration" and therefore - to the limit - a point event (x, y, z, t).

---

[1] I showed in my book that the mysterious expression "law of change of states" Einstein's principle of relativity could be the Boltzmann law. We notice that k is at the level of a principle theory while h is only a the level of a constructive theory (see conclusion). This is the reason for which I have notice N/R by k.

By an elementary event we will understand an event that is supposed to be concentrated in one point and is of infinitely short duration."[Einstein, SR, 1910] "To determine the position of a process of infinitesimally short duration that occurs in a space element (point event) [Einstein, SR, 1907]

We have so shown that the concept of independent events is the same that the concept of independently moving "points". Each independent state of a moving "point" is not characterised by **(x, v)** - as in the Gibbs mechanical statistics- but by (x, y, z, t). We have however not yet shown the connection with Einstein's independently quanta of light. Let us go on with Einstein's quotation:

Paragraph 6 "Interpretation according to Boltzmann's principle of the expression for the dependence of the entropy of monochromatic radiation on volume

In section 4 (limiting law for the entropy of monochromatic radiation at low radiation density) for the dependence of the entropy of monochromatic radiation on volume (…)

$$S - S_0 = k \ln(\frac{V}{V_0})^{\frac{1}{\beta k}\frac{E}{\nu}}$$

And compare to the general formula expressing the Boltzmann's principle

$$S - S_0 = k \ln W$$

We arrive in the following conclusion.

If monochromatic radiation of frequency ν and energy E is enclosed (by reflecting walls) in the volume V, the probability that at a randomly instant the total radiation energy will be found in the portion V of the volume $V_0$ is:

$$W = (\frac{V}{V_0})^{\frac{1}{\beta k}\frac{E}{\nu}}$$

From this we further conclude that monochromatic radiation of low density behaves thermodynamically as if it consisted of **mutually independent energy quanta** of magnitude hν.

Einstein obtains in his paragraph 4 a relation between entropy and volume for the Wien radiation and in his paragraph 5 a similar relation thanks to the applying of Boltzmann's principle as the fundamental law of movement for his independent points (W-> S). Einstein's strategy in paragraph 6 consists logically in a comparison and an inversion (S->W).

The famous disagreements between Planck and Einstein about the structure of light and the definition of probability are the same thing. Einstein's inversion of Boltzmann principle means not only that the relation may be mathematically in the other sense but implies *physically* that "the probability that at a randomly instant the total radiation energy will be found in the portion V" in others words, if Einstein's probability has a physical sense, Einstein's quanta has also a physical sense.

Einstein's "points" can be as well material point or light point whose the singularity consists in his velocity v = c. The deepest meaning of the relation E = h ν in Einstein's paper is that the structure of monochromatic radiation consists in a discontinuous series of independent point events.

## 4-Events, points of light and complex of light in June paper

It is well known that Einstein's kinematics implies the deletion of the ether. The first meaning of the existence of this medium is that the light is a wave i.e. "there is no point with velocity of light (v = c)".

In the first part of his article, Einstein never speaks about "wave of light". Einstein's basic concept is "ray of light" (in the sense of geometrical optics), i.e. a point of light. In his paragraph 3, Einstein deduces the transformation of the co-ordinates of an event:

To any system of values x, y, z, t, which completely defines the place and the time on an *event* in the stationary system K, there belongs a system of values ξ, η, ζ, τ, determining that *event* relatively to the system k, and our task is now to find the system of equations connecting this quantities (…). We obtain:

$$x = \varphi \gamma (x - vt), \eta = \varphi y, \zeta = \varphi z,$$
$$\tau = \varphi \gamma (t - vx/c^2)$$

We must show now that Einstein's transformations of the co-ordinates of an event supposes an extension of the classical notion of material point. In other words: it is not only in the first part of his article that Einstein characterises the light as a point but also in the second part.

In the famous polemic of priorities (see conclusion) between Poincaré and Einstein, Whittaker, who attributes the discovery of SR to Lorentz and Poincaré, conceded that the Doppler formula was deduced by Einstein and not by Poincaré. Is it a historical contingence?

The first time that Einstein speaks of "the electrodynamic (sic) wave", in his famous June paper, he applies his Lorentz's transformation at … a point v = c of the plane of a wave plane (paragraph 7):

In the system K and very far from the coordinate origin, let there be a source of electrodynamic waves …[whose the phase is] $\Phi = \omega t - \mathbf{k} \mathbf{x}$. We want to known the character of these waves when investigated by an observer at rest in the moving system k. Applying the transformation equations for electric and magnetic forces found in section 6 and those for coordinates in section 3, we immediately obtain: $\Phi' = \omega \tau - \mathbf{\kappa} \mathbf{\xi}$.

It is of course forbidden to make v = c in the Lorentz transformations (LT) but the state (x, y, z, t) of a light point is an event. The propagation of the light consists in Einstein's SR fundamentally in a series of events. The concept of event implies a extension of the domain of application of LT.

The physicists credit often Einstein with the discovery of the relativistic law of composition of the speed and Poincaré with the discovery of the structure of group of Lorentz transformations (LT). But in fact the two elements are in each approach (Einstein's § 5 and Poincaré's § 4). The interesting point is not in these polemical questions of priority but the fact that Einstein applies his LT for light point unlike Poincaré whose the law of composition of velocities concerns the material systems or the material points:

It is noteworthy that the Lorentz transformations form a group. For, if we put:

$$x' = l k (x + \varepsilon t), \quad y' = l y, \quad z' = l z, \quad t' = k l (t + \varepsilon x)$$

and

$$x'' = l' k'(x' + \varepsilon' t'), \quad y' = l' y, \quad z' = l' z, \quad t'' = l' k' (t' + \varepsilon' x')$$

we find that

$$x' = l'' k'' (x + \varepsilon'' t), \quad y'' = l'' y, \quad z'' = l'' z, \quad t'' = l'' k'' (t + \varepsilon'' x)$$

with

$$\varepsilon'' = \frac{\varepsilon + \varepsilon'}{1 + \varepsilon\varepsilon'}$$

[2d, §4, 1905]

Einstein introduces his kinematics theorem of addition of velocities in his paragraph 5:

If w also has the direction of the x-axis v, we get

$$u = \frac{v + w}{1 + \frac{vw}{c^2}}$$

It follows from this equation that the composition of two velocities that are smaller than c always results in a velocity that is smaller than c.
It also follows that the velocity of light c cannot be changed by compounding it with a "subliminal velocity". For this case we get:

$$u = \frac{c + w}{1 + \frac{w}{c}} = c$$

Poincaré's relation is formally the same than Einstein's one but it concerns only the material systems or material points. The most important difference between the two conceptions consist in the case v = c. Einstein The reader could be think that Einstein's substitution v = c is not fundamental but Einstein writes in his paragraph 4:

For v = c, all moving objects – considered from the "rest" system, - shrink into plane structures. For superluminal velocities our considerations become meaningless.

We return now in the second part of the article. In his paragraph 8, Einstein deduced, after the Doppler relations, the way of relativistic transformation of energy and frequency of what he calls a "complex of light".

It is noteworthy that the energy and the frequency of a light complex vary with the observer's state of motion according to the same law.

Many authors have of course underlined the similarity between quanta of light of January and complex of light of June. They generally said that Einstein's deduction of proportionality relation between frequency and energy confirms only that Einstein's SR is a theory of principles and therefore not depending of a classical or quantum model of the light.
It is necessary to say that but *it is not sufficient* because to obtain this relation energy-frequency, Einstein has one more time applies the Lorentz transformation to a light point v = c (paragraph 8).

The surface elements of the spherical surface [in K] moving with the velocity of light are not traversed by any energy. We may therefore say that this surface permanently encloses the same light complex. We investigate the quantity of energy from the system k, i.e. the energy of the light complex relative to the system k.

In order to obtain his quantum relation, Einstein considers a quite singular limit case not only for the applying of LT but also for Maxwell-Lorentz equations themselves. Lorentz, who doesn't appreciate this approach of the young Einstein writes in the final proceedings of the Conference in Brussels (1911):

There is nothing in Maxwell equation that can encloses a quantity of energy in a limited volume

The light complex were not more appreciated by the other physicists. The young Einstein was however not wrong because the volume and the energy changes in function of the time. In fact, Einstein treats, in a very singular manner, the Maxwell equations not as an wave equation but as equation of movement of a light point. Light complex is a finite volume around the light point characterised by a frequency. We find again to the limit the concept of event and that Einstein's representation of the light is deeply discontinuous.

Penrose writes:

To me it is virtually inconceivable that he would have put forward two papers in the same year which depended upon hypothetical views of nature that he felt were in contradiction with each other.[1. preface]

We proved therefore that is no contradiction: quanta of light and complex of light are on the same side of the quantum-classic frontier, the most important of the present physics. When the young Einstein considers the Maxwell equations as an application his kinematics of events, he doesn't treat them classically.

His light complex associate to a point light wavy properties (the first being already that the light point is a point whose the speed is independent of the source that emits it). We find again "de Broglie" but not for the electron but for the photon, in the strong quantum sense of the concept.

However, we don't want say here that SR, in the standard sense where SR is the group properties of LT, imposes a quantum conception of the light. Indeed we must now prove that there exist another SR based on a purely wavy representation of the light and on a implicit kinematics of material point.

**5- Poincaré's principles of implicit relativistic kinematics of deformable rods**

The main mistake of Poincaré's supporters in the discover of SR (1905) is to extract elements in Poincaré's work (for example, the synchronisation of the clocks, the scepticism with respect to the ether, the statement of principle of relativity, the structure of group of LT …) and to insert them in Einstein's logic (or axiomatic) in order to show that Poincaré had "Einstein before Einstein".

This is of course unpleasant for Einstein but in my opinion also for Poincaré.

Indeed with this method of extraction-insertion it is absolutely impossible to understand Poincaré's relativistic logic. Poincaré's logic is *deeply* different than Einstein's one because his theory of relativity proves the existence of the ether while Einstein's one proves the absence of this latter: there exists a genuine antinomy in the sense of Kant.

I show in my book [2] that Poincaré's scepticism dates back to long before his crucial interest for Lorentz theory based on the existence of the ether and the electrons.

According to Poincaré the main problem of Lorentz theory is his incompatibility with the principle of reaction (Poincaré 1900): what is the dynamical force for responsible of the real contraction of length in ether (Lorentz hypothesis, LH)?

Poincaré's guide for the theory of relativity is clearly the third principle of mechanics in the same epistemological manner that the second principle of thermodynamics is Einstein's guide. In his fundamentally 1905 work he determines this force exerted by the ether on the electron:

But in the Lorentz hypothesis [LH], also, the agreement between the formulas does not occur just by itself; it is obtained together with a possible explanation of the compression of the electron under the assumption that the *deformed and compressed electron is subject to constant external pressure,* the work done by which is proportional to the variation of volume of this electron." (my italics) [2d, intro.]

In his 1905 work Poincaré shows *dynamically* (with his pressure of ether) the compatibility between the principle of relativity and principle of real contraction:

So Lorentz hypothesis [LH] is the only one that is compatible with the impossibility of demonstrating the absolute motion [RP]; if we admit this impossibility, we must admit that moving electrons are contracted such a manner to become revolution ellipsoids whose two axis remain constant [2d, § 7].

It is true that there is no theory of space-time in Poincaré's 1905 work. There is however a implicit kinematics (real contraction of deformable rod) behind his dynamics (Poincaré deduce fundamental equations of relativistic dynamics by introducing the ether pressure).

Implicit Poincaré's kinematics is based on two postulates (principle of relativity and LH) exactly of the same manner that explicit Einstein's kinematics is based on two postulates (principle of relativity and light principle).

*Poincaré's SR with ether implies another use of LT:*

In accordance with LH, moving electrons are deformed in such a manner that the real electron becomes an ellipsoid, while the ideal electron at rest is always a sphere of radius r (…) The LT replaces thus a moving real electron by a motionless ideal electron". (2d, §6)

In order to illustrate this, we can use Tonnelat's diagram [5]

(fig 1, *we adopt Poincaré's and Einstein's respective notations* in the following, see former respective quotations about LT):

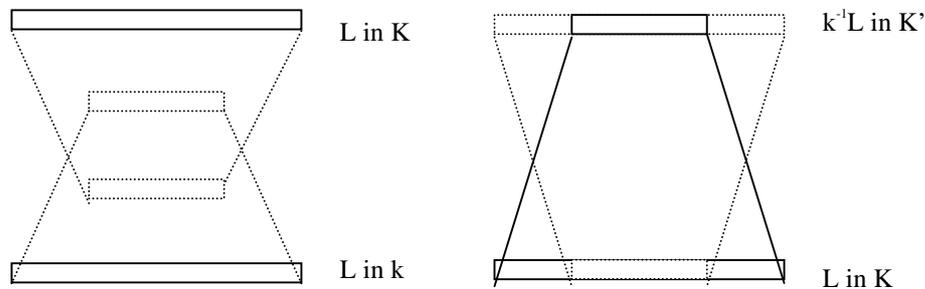

**Fig 1**
In Einstein's words
(dashed lines):
"The length to be discovered [with LT] "the length of the (moving) rod in the stationary system"

**Fig 2**
In Poincaré's words
(dashed lines):
"LT replaces thus a moving real electron [rod] by a motionless ideal electron [rod]"

In **fig 1**, in Einstein's standard SR, the contraction of the moving rod $\gamma^{-1}L$ is not real (dashed line) but is the reciprocal result of a comparison of measurement made on identical rods[2] L (continuous lines) from one system to the other with the well known use of LT (dashed lines).

---

[2] Max Born, who was a specialist of rigidity in Einstein's special relativity: "A fixed rod that is at rest in the system S and is of length 1 cm, will, of course, also have the length 1 cm, when it is at rest in the system S', provided that the remaining physical conditions are the same in S' as in S. Exactly the same would be postulated of the clocks. We may call this *tacit assumption* of Einstein's theory the *"principle*

In **fig2**, in Poincaré's SR the contraction of the moving rod $k^{-1}L$ is *by principle* (LH) real (continuous lines) in K'. By the use of LT the length of the rod in K' (for observers in K') seems be equal to L (dashed lines). Reciprocally, we can of course reverse the role of K and K'(where ether is now chosen at rest) and reverse the continuous lines and dashed lines.

The calculation with the LT is also very easy. Suppose the ether is chosen by definition at rest in K. The real length of the rod placed in the moving system K' is thus $k^{-1}L$. The first LT $x' = k(x - \varepsilon t)$ "replaces" (in Poincaré's terms) in any time t (see below) the length moving real rod $k^{-1}L$ by a motionless rod L.

According to Poincaré, the real differences are *compensated* by a "good use" of LT. According to Einstein, the *identical* processes seem to be different by another "good use" of LT. This is of course valid for the rods but also for the clocks.

Now we must show that Poincaré's LT is fundamentally a transformation of the coordinates of a material point and not a transformation of the co-ordinate of an event.

**5- Independent true time, material points and wave of light in Poincaré's SR with ether**

We showed that the ether plays an essential role in Poincaré's SR. But it is not an absolute ether in the sense of Lorentz. Indeed, according to Poincaré only a *relative* speed can be measured with respect to the ether. The group structure implies that Poincaré's relativistic ether can be chosen, for all the couples of inertial systems, at rest in one of the two frames but the other is then in movement with respect to the ether. The relativistic ether supposes that the light is a wave.

The direct consequence of wavy representation of the light is Poincaré's convention of synchronisation based on the duality local time and true time. Indeed the two inertial systems are not in the same state of movement with respect to the relativistic ether. In his talk on "The principles of mathematical physics", Poincaré considers optical signals as optical perturbations:

**1-** (Two observers are at rest relative to ether, System K)

"The most ingenious idea has been that of local time. Imagine two observers [A and B] who wish to adjust their watches by optical signals; they exchange signals, (…) And in fact, they [The clocks of A and B] mark the same hour at the same physical instant, but on one condition, namely, that the stations are fixed.

**2-** (Two observers are moving relative to ether, System K')
In the contrary case the duration of the transmission will not be the same in the two senses, since the station A, for example, moves forward to meet the optical perturbation emanating from B, while the station B flies away before the perturbation emanating from A."
The watches adjusted in that manner do not mark, therefore the *true time*; they mark the *local time*, so that one of them goes slow on the other (de telle manière que l'une retarde sur l'autre). It matters little, since we have no means of perceiving it." (…)
(for exact compensation, we must add LH, former quotation)
Unhappily, that does not suffice, and complementary hypotheses are necessary. It is necessary to admit that bodies in motion undergo a uniform contraction in the sense of the motion." [2b]

Poincaré's synchronisation is clearly based for his second system K' on the *duality of the true time t and the local time t'*.

---

*of the physical identity of the units of measure"."* [4, my italics p252] This is not really a third hypothesis because Einstein's deduces the identity of his rods from his relativity principle. The rigidity is not important. The important thing in the spirit of the young Einstein's text, is to postulate the existence in Nature of processes giving units of length and time.

This is Poincaré's "tour de force" to have shown that his second principle (LH) implies for the local time – not the Lorentz expression t' = t + εx - but the expression given by the fourth LT: t' = k (t + εx).

We proved[3] that Einstein's radical elimination of ether implies that he prepares identically his two systems K(t) and k(τ), ("light clocks" or better "clocks with photons") in the same state of synchronisation.

This is the reason for which the duality true time-local time have no sense in Einstein's logic. Poincaré's relativistic ether hasn't got a particular state of movement, but *his two systems are never in the same state of moving relative to it*. This is the direct consequence of the wavy nature of the light that propagates in a medium. Poincaré's local time is not a *internal time, given by identical clocks,* in the second frame K'.

The main interest here is the incompatibility of the concept of local time and the concept of event. The local time $t'_A = k(t + \varepsilon x_A)$ (or the apparent time in the vocabulary of Poincaré) is a time that depends of three factors: firstly of course the independent true time (independent doesn't mean absolute), secondly of the speed (the difference of state of moving of the two systems relative to ether) and thirdly of the place of the observer placed for example in A:

Let us consider one point at true time with the true coordinates x, y, z. What will be the apparent coordinates x', y', z' at apparent time t'?

$$x' = k(x + \varepsilon t), \qquad y' = y, \qquad z' = z, \qquad t' = k(t + \varepsilon x)$$

Let us the units by such a way that the speed of light equals to 1. What is the meaning of ε? The translation velocity of the system in the sense of the x-axis et has the value –ε.
In the same way t' is the apparent time, *because in two points*, the apparent time differs of the true time with a quantity proportional to the abscissa. [Cours a la Sorbonne, 3a]

The local time (apparent time) in A means that the time of an "event"[4] in B is not the time given at the place B where this event occurs. This is not a time of event in Einstein's sense of the coincidence of two independent events. It is not necessary to consider the distance simultaneity (necessarily independent events) in both SR because already Einstein's definition of simultaneity (coincidence) at the same place is sufficient.

Even if Poincaré doesn't use the concept of event, let us introduce an "event" in weak sense i.e. in the sense of a point on the trajectory of material point. We must then of course have two material points to consider the simultaneity of two "events". We underline then that in the well known *laplacian deterministic point of view* of the classical mechanics the two "events" considered (for example the two points are simultaneously at a given distance) cannot be independent. This is exactly what Poincaré admitted in his text on "L'espace et le temps" en 1911 [3a].

The choice between the independence of (true) time (Poincaré) and the independence of events (Einstein) is then clearly antinomic.

**Conclusion**

The fact that SR is basically considered as a theory of space-time implies that historically Einstein must be considered as the author of this theory.

---

[3] Einstein's synchronisation of identical clocks within the second system k is exactly the same than the synchronisation in the first system K because the speed of light is of course exactly the same. "It is essential to have time defined by means of stationary clocks in stationary system (…)" [ §1, June 1905]. To do this [deduce LT] we have to express in equations that τ is nothing else than the set of data of clocks at rest in system k, which have been synchronized according to the rule given in paragraph 1".

[4] Even if Poincaré doesn't use concept of event let us introduce an "event" in weak sense i.e. in the sense of a point on the trajectory of material point. We must then of course have two material points to consider the simultaneity of two "events". We underlines then that in the laplacian point of view of the classical mechanics the two "events" considered (for example the two points are simultaneously at a given distance) are not independent.

Indeed Poincaré has not developed in 1905 as Einstein a kinematics based on two principles but "only" a relativistic dynamics or more precisely a relativistic dynamic of continuous medium. But the opposition between SR "with principles" (Einstein) and SR "without principles" (Poincaré) is superficial (historical contingence) because it is easy to see that there is a implicit kinematics of material point and wave of light behind Poincaré's relativistic dynamics.

The comparison of Einstein's kinematics of events with Poincaré's implicit kinematics of material point is very interesting because it shows the deep and irreducible singularity not only about Einstein's principle of identity of units of measure but also about Einstein's "points of light", "quanta of light", "complex of light" or "photons".

The idea of a "fine structure" of SR means that the borderline classical-quantum, the main cut of the physics of XX century, passes between the two SR.

The reader could think that we contest that Einstein's SR is based on principles because we show that it exist a quantum representation hidden behind this principles. But quantum theory is no longer a constructive or an heuristic theory as at the epoch of old quantum theory. The quantum theory is now considered as a theory of principles or a genuine picture of world.

The existence of a "fine structure" of SR opens a new way of research that consists to transform Einstein's SR – *without any change in his historical originality* - under the "enormous pressure" of classical Poincaré's theory, CSR) in a real relativistic quantum theory (QSR) of the light (photons), the time (quanta of time) and the space (acausal zone).

In others words, we can overtake the fierce concurrency between Einstein and Poincaré on the market of the priorities in a genuine dialectical interaction between the two complete and coherent relativistic logics. The philosophical opposition "continous-discontinous" becomes then a physical opposition "classical-quantum".